# Uncertainty Quantification in density estimation from Background Oriented Schlieren (BOS) measurements

Lalit K. Rajendran[1], Jiacheng Zhang[2], Sayantan Bhattacharya[2], Sally P. M. Bane[1] and Pavlos P. Vlachos[2]*

[1]Purdue University, School of Aeronautics and Astronautics, West Lafayette, USA.

[2] Purdue University, School of Mechanical Engineering, West Lafayette, USA.

*pvlachos@purdue.edu

## Abstract

We present an uncertainty quantification methodology for density estimation from Background Oriented Schlieren (BOS) measurements, in order to provide local, instantaneous, a-posteriori uncertainty bounds on each density measurement in the field of view. Displacement uncertainty quantification algorithms from cross-correlation based Particle Image Velocimetry (PIV) are used to estimate the uncertainty in the dot pattern displacements obtained from cross-correlation for BOS and assess their feasibility. In order to propagate the displacement uncertainty through the density integration procedure, we also develop a novel methodology via the Poisson solver using sparse linear operators. Testing the method using synthetic images of a Gaussian density field showed agreement between the propagated density uncertainties and the true uncertainty. Subsequently the methodology is experimentally demonstrated for supersonic flow over a wedge, showing that regions with sharp changes in density lead to an increase in density uncertainty throughout the field of view, even in regions without these sharp changes. The uncertainty propagation is influenced by the density integration scheme, and for the Poisson solver the density uncertainty increases monotonically on moving away from the regions where the Dirichlet boundary conditions are specified.

# Nomenclature

| | | | |
|---|---|---|---|
| $u$ | Measurement | $K$ | Gladstone-Dale constant |
| $\delta_u$ | Measurement Error | $\nabla$ | Gradient |
| $\sigma_u$ | Measurement Uncertainty | $S$ | Source Term |
| $u_0$ | True Value | $\bar{\bar{\nabla}}$ | Mapping matrix |
| $\sigma$ | Standard Deviation | $\bar{\bar{L}}$ | Label matrix |
| $t_{CI}$ | Coverage Factor | $\bar{\bar{\Delta}}$ | Laplacian operator matrix |
| $\rho$ | Density | $\bar{\bar{\Sigma}}$ | Covariance matrix |
| $\rho_p$ | Projected Density | $X_0, Y_0$ | Centroid |
| $\Delta\vec{x}$ | In-plane Displacement | $E$ | Expectation operator |
| $M$ | Magnification | $n$ | Refractive index |
| $Z_D$ | Distance between dot pattern and density gradient field | | |

## 1. Introduction

Background Oriented Schlieren (BOS) is a flow measurement technique, where the apparent distortion of a dot pattern viewed through a medium with refractive index gradients is measured using cross-correlation, tracking or optical flow based algorithms to estimate the density gradients in the medium [1]–[4]. The density gradients can be integrated spatially to obtain the density field, generally by solving the Poisson equation using different computational procedures [5]. Owing to the simple setup and ease of use, BOS has been applied widely in laboratory scale as well as in large scale and rugged industrial facilities, and is becoming the preferred method of density measurement in fluid flows [6]–[13].

BOS measurements are increasingly used for Computational Fluid Dynamics (CFD) model validation and design [10], [12], [14]–[16]. However, currently there is no framework for quantifying the uncertainties in the density estimation, and inform proper validation of computational models. The BOS measurement chain is complex and subject to several sources of uncertainties ranging from the dot pattern parameters (dot size, dot density), non-uniform illumination, vibrations, blurring/out-of-focus effects, non-linearities and small scale fluctuations in the flow field, uncertainties in measurement of the optical layout, as well as the processing and post processing methodologies used to calculate the density from the image displacements. As a result, the uncertainty on the final density measurement is a high-dimensional, coupled, non-linear and non-trivial function of several parameters, and can vary widely across the field of view and across a time series of measurements. Therefore, a comprehensive method for estimating and reporting uncertainties on BOS density measurements is needed. This paper aims to develop and test the first uncertainty quantification methodology to provide a-posteriori, local, instantaneous uncertainty bounds for each density measurement in the field of view for a BOS experiment.

For a measurement $u$, the uncertainty $\sigma_u$ is defined as the interval around the measurement in which the true value $u_0$, and by extension the true error $\delta_{u_0}$, is believed to exist with a predetermined degree of confidence [17]. Following ISO-GUM [18], the standard uncertainty is defined as the range of measurement values that are one standard deviation $\sigma$ about the true value, for an arbitrary parent population. The expanded uncertainty is defined for an assumed parent distribution for the error, and is specified using a confidence interval at a defined percentage and a coverage factor $t_{CI}$. This is to indicate that the true value/error lies in an interval $\sigma_u = t_{CI}\sigma$ around the measurement for the pre-defined percentage of samples drawn from the parent distribution. For example, if the errors are drawn from a Gaussian distribution, the expanded uncertainty at 68% confidence interval is equal to the standard uncertainty ($\sigma_u = \sigma$), and the expanded uncertainty at 95% confidence interval is equal to approximately twice the standard uncertainty ($\sigma_u = 1.96\sigma$). The standard uncertainty is reported throughout this paper.

In the related field of Particle Image Velocimetry (PIV) [19]–[22], there have been widespread efforts in the past decade to develop a-posteriori uncertainty quantification methodologies [23], [24], as well as to perform comparative assessment of the existing methods [24]–[26]. As the displacement estimation in BOS is similar to PIV, PIV-based displacement uncertainty methods can be applicable to BOS measurements.

Displacement uncertainty estimation methods for 2D planar PIV can be broadly divided into indirect and direct methods. Indirect methods predict the displacement uncertainty by calibrating the variation of uncertainty to various image parameters and signal to noise ratio metrics, where the calibration is obtained using Monte-Carlo simulations with synthetic images. Timmins et. al. proposed the first PIV uncertainty quantification method termed 'Uncertainty Surface' (US) [27], where the uncertainty is calibrated based on four particle diameter, seeding density, displacement and shear. Charonko and Vlachos [28] proposed the Peak to Peak Ratio (PPR) method, where the uncertainty is calibrated against and calculated using the ratio of the primary to secondary cross-correlation peak heights. This method was later generalized by Xue et. al. [29] to other correlation plane derived metrics such as the Peak to Root Mean Square Ratio (PRMSR), Peak to Correlation energy (PCE), and cross-correlation. The Mutual Information (MI) based uncertainty quantification by Xue et. al. [30], defined as the effective number of correlated particle pairs between two image frames, was also used to estimate PIV uncertainty. The performance of all indirect methods relies on the calibration process, which must be accurate and reflect all possible experimental scenarios in a typical measurement.

On the other hand, direct methods estimate uncertainty directly from the properties of the image or correlation plane and do not require any calibration. Examples of direct uncertainty estimation methods include the Image Matching (IM) method proposed by Sciacchitano et. al. [31], Correlation Statistics (CS) method proposed by Wieneke [32] and the Moment of Correlation (MC) proposed by Bhattacharya et. al. [33]. Each of the direct methods has a different working principle, and in the following, we briefly describe the assumptions, working principles and limitations of the direct methods.

Image Matching (IM) or Particle Disparity (PD) proposed by Sciacchitano et. al. [31], estimates the uncertainty in the displacement using a statistical analysis of the disparity between the measured positions of particles or dots in the two frames after a converged iterative deformation interrogation procedure [34], [35]. This method requires at least six particles to be in the interrogation window for statistical calculations but fails at high seeding densities due to errors in particle identification. This method is also affected by image noise and loss of particles between frames, especially due to out of plane motion [25]. Correlation Statistics (CS) proposed by Wieneke [32], estimates the uncertainty again using the image disparity but at a pixel level. The asymmetry of the correlation peak at the end of a converged window deformation procedure is used as measure of the correlation error and the standard deviation of the error is propagated through the subpixel estimator to estimate the displacement uncertainty. As the method relies on statistics of the correlation plane, it works better with higher seeding densities and larger interrogation windows [25], [32]. Moment of Correlation (MC) proposed by Bhattacharya et. al. [33] predicts the uncertainty by estimating the second order moment of the cross-correlation plane. The estimation process involves the calculation of the Generalized Cross Correlation (GCC) from the inverse Fourier transform of the phase of the complex cross-correlation plane [36]–[38]. The primary peak region of the GCC plane represents the probability density function (PDF) of all possible displacements for the given interrogation window [33]. This PDF is convolved with a Gaussian function, corrected for peak broadening by displacement gradients, and normalized by the effective number of correlating pixels (calculated using MI [30]) to estimate the uncertainty. Similar to Correlation Statistics, this method also works better with high seeding densities and large interrogation windows, as small interrogation windows can lead to an over-prediction of the uncertainty [33].

In two independent comparative assessment of the methods, Sciacchitano et. al. [25] and Boomsma et. al. [26] found the direct methods to be more sensitive to variations in the random error, though Boomsma et. al. [26] found the direct methods to underpredict the standard uncertainty in some cases. Since the indirect methods rely on calibration, only the direct methods will be considered in this work.

There have also been efforts to propagate the displacement uncertainties in PIV derived quantities. Wilson and Smith [39] extended the displacement uncertainties calculated from the Uncertainty Surface method to estimate uncertainties in mean and fluctuating velocity statistics. Sciacchitano and Wieneke [40] provided a framework for calculation of uncertainties for displacement gradient based quantities such as the vorticity, and also identified the importance of spatial correlation of the displacement errors. Bhattacharya et. al. [41] proposed a methodology for stereo-PIV uncertainty quantification by accounting for the uncertainties introduced in the calibration and self-calibration process, along with the planar correlation uncertainty for individual camera image correlation. Azilji et. al. [42] proposed a methodology based on a Bayesian framework to calculate the uncertainties for PIV-based pressure measurement in a three-dimensional flow field, though they calculated the displacement uncertainty from the divergence error of the velocity field and did not use any of the above mentioned displacement uncertainty quantification methods.

In this paper we propose and implement the first comprehensive framework to model and propagate uncertainties from displacement measurements in a BOS experiment onto the final density measurement. To do this, we use methodologies for PIV uncertainty quantification [26]–[28], [43]–[48] and propagate these uncertainties through the BOS measurement chain including the density gradient integration and density reconstruction. We test both the PIV displacement uncertainty schemes as well as the uncertainty propagation framework with synthetic and experimental BOS images.

## 2. Methodology

The proposed uncertainty quantification methodology closely follows the BOS measurement chain and is illustrated in Figure 1. First, the raw image pairs and processed displacement fields are used along with PIV-based uncertainty estimation methods to calculate the local, instantaneous uncertainty on each displacement vector as indicated in Figure 1 (a). Following this, the optical system parameters such as the magnification and the distance between the dot pattern and the density gradients are used to estimate the uncertainty in the projected density gradient field.

In BOS experiments, the projected density gradient field is related to the apparent displacement of the dot pattern by

$$\nabla \rho_p = \int \nabla \rho \, dz = \frac{\Delta \vec{x}}{Z_D M} \frac{n_0}{K} \qquad (1),$$

where $\Delta \vec{x}$ is the displacement, $M$ is the magnification of the dot pattern, $Z_D$ is the distance between the dot pattern and the mid-point of the density gradient field, $n_0$ is the ambient refractive index, $K$ is the Gladstone-Dale constant (= 0.225 cm³/g for air) and $\rho_p = \int \rho \, dz$ is the projected density field.

Similarly, the uncertainty in the projected density gradient field can be expressed by

$$\sigma_{\nabla \rho_p} = \frac{\sigma_{\Delta \vec{x}}}{Z_D M} \frac{n_0}{K} \qquad (2),$$

where $\sigma_{\Delta \vec{x}}$ is the displacement uncertainty and $\sigma_{\nabla \rho_p}$ is the uncertainty in the projected density gradient field (Figure 1 (b)). It should be noted that some of the experimental parameters occurring in the above equations can also have their own uncertainties such as the magnification $M$ and the distance $Z_D$. However, herein, for simplicity we will assume these as known and constant, as our focus is on propagating the displacement-based uncertainties. Any uncertainties in these quantities can be handled in a straightforward manner using the Taylor series propagation model [17].

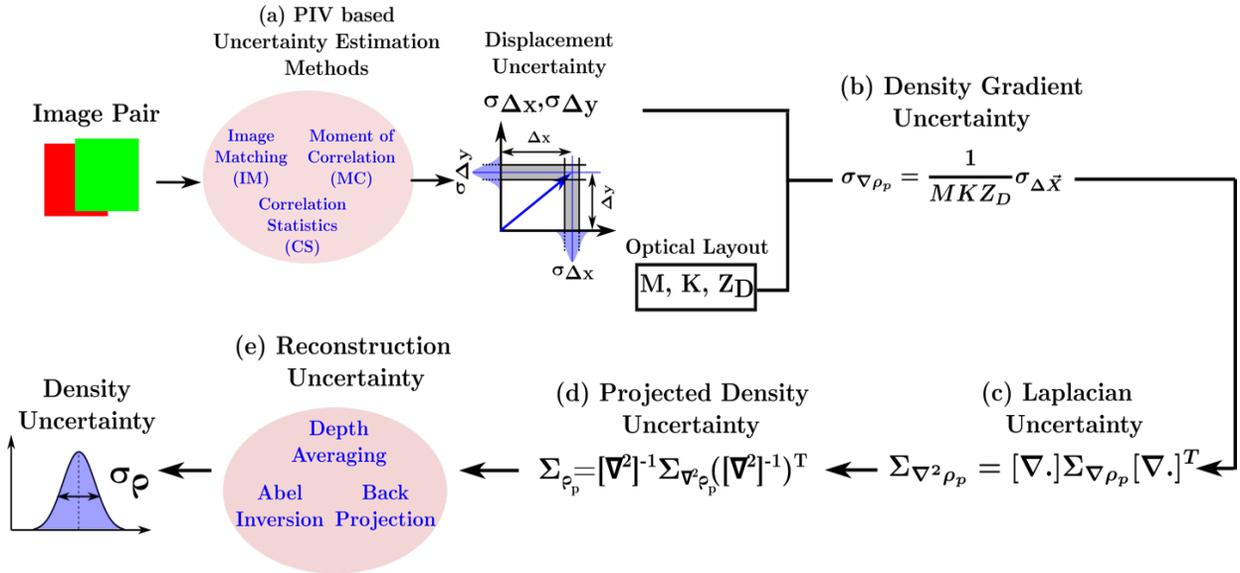

**Figure 1.** Proposed uncertainty quantification methodology for BOS measurements.

The next step in BOS experiments is to calculate the projected density field by solving the Poisson equation:

$$\frac{\partial^2 \rho_p}{\partial x^2} + \frac{\partial^2 \rho_p}{\partial y^2} = S \qquad (3)$$

where the source term $S$ denotes the Laplacian of the density field calculated from the projected density gradient field. This equation is then discretized into a system of linear equations using finite difference schemes and solved using appropriate boundary conditions (Dirichlet/Neumann) depending on prior knowledge about regions of the flow field.

The discretization and solution procedure are as follows. The source term $S$ is calculated as

$$\bar{S} = \bar{\bar{\nabla}}_x \overline{\frac{\partial \rho_p}{\partial x}} + \bar{\bar{\nabla}}_y \overline{\frac{\partial \rho_p}{\partial y}} \quad (4),$$

where $\bar{\bar{\nabla}}_x, \bar{\bar{\nabla}}_y$ are the discretized gradient operators (matrices represented by a double overbar) that depend on the finite difference scheme, and $\overline{\frac{\partial \rho_p}{\partial x}}, \overline{\frac{\partial \rho_p}{\partial y}}$ are the density gradients (1D column vector represented by a single overbar). A second order central difference discretization scheme is used for the results reported in this paper.

The source term is combined with points on the boundary to create an augmented matrix $\bar{\bar{R}}$, given by

$$\bar{\bar{R}} = \bar{\bar{\nabla}}_x \overline{\frac{\partial \rho_p}{\partial x}} + \bar{\bar{\nabla}}_y \overline{\frac{\partial \rho_p}{\partial y}} + \frac{1}{h^2} \bar{\bar{L}} \bar{\rho}_{p,L} \quad (5),$$
$$= \bar{\bar{S}} + \bar{\bar{S}}_L$$

where $\bar{L}$ is a label matrix specifying points on the boundary, and $h$ is the grid spacing, and $\bar{\rho}_{p,L}$ is the array of densities of the points corresponding to the label matrix (the Dirichlet boundary condition).

The projected density is calculated by multiplying the augmented matrix $\bar{\bar{R}}$ with the inverse of the augmented Laplacian operator

$$\rho_p = \begin{bmatrix} \bar{\bar{\Delta}} & 0 \\ 0 & \frac{\bar{\bar{L}}}{h^2} \end{bmatrix}^{-1} \begin{bmatrix} \bar{\bar{S}} \\ \bar{\bar{S}}_L \end{bmatrix} \quad (6),$$

where $\bar{\bar{\Delta}}$ (sometimes also represented by $\nabla^2$) is the discretized Laplacian operator for the interior points corresponding to the finite difference schemes used to calculate the Laplacian ($\bar{\bar{\Delta}} = \bar{\bar{\nabla}}_x \bar{\bar{\nabla}}_x^T + \bar{\bar{\nabla}}_y \bar{\bar{\nabla}}_y^T$). Thus, equation (6) essentially solves the Poisson equation (3) to give the projected density field $\rho_p$.

The uncertainty calculations are performed in a manner similar to the density integration, by propagating the covariances through the finite difference operators and accounting for the boundary conditions used to calculate the corresponding density field. The covariance in the augmented source term defined in equation (5) is given by

$$\bar{\bar{\Sigma}}_R = \bar{\bar{\nabla}}_x \bar{\bar{\Sigma}}_{\frac{\partial \rho_p}{\partial x}} \bar{\bar{\nabla}}_x^T + \bar{\bar{\nabla}}_y \bar{\bar{\Sigma}}_{\frac{\partial \rho_p}{\partial y}} \bar{\bar{\nabla}}_y^T + \frac{1}{h^2} \bar{\bar{L}} \bar{\bar{\Sigma}}_{\rho_{p_L}} \frac{1}{h^2} \bar{\bar{L}}^T \quad (7),$$
$$= \bar{\bar{\Sigma}}_S + \bar{\bar{\Sigma}}_{S_L}$$

and as shown in Figure 1 (c), the covariance in the projected density ($\bar{\bar{\Sigma}}_{\rho_p}$) is calculated using

$$\bar{\bar{\Sigma}}_{\rho_p} = \begin{bmatrix} \bar{\bar{A}} & 0 \\ 0 & \frac{\bar{\bar{L}}}{h^2} \end{bmatrix}^{-1} \begin{bmatrix} \bar{\bar{\Sigma}}_S \\ \bar{\bar{\Sigma}}_{S_L} \end{bmatrix} \left( \begin{bmatrix} \bar{\bar{A}} & 0 \\ 0 & \frac{\bar{\bar{L}}}{h^2} \end{bmatrix}^{-1} \right)^T \quad (8).$$

Finally, the uncertainty in the density is calculated from the square root of the diagonal terms of the density covariance matrix as indicated in Figure 1 (d) and is expressed as,

$$\sigma_{\rho_p} = \sqrt{\text{diag}(\Sigma_{\rho_p})} \quad (9).$$

All linear operators in the solution procedure are modeled as sparse matrices to increase computational speed.

The next step is to calculate the 2D density field from the projected density field, either by depth averaging (dividing the projected density field by the thickness of the density gradient field) if the extent of the density field is known, or through an Abel inversion or Filtered Back Projection (FBP) procedure if the flow field is axisymmetric (Figure 1 (e)) [7]. While each reconstruction procedure can create a different amplification of the uncertainty, only the depth averaged reconstruction approach will be considered in this paper. For situations which involve the use of Abel inversion, the uncertainty can again be propagated through a matrix representation of the Abel inversion procedure, because all the Abel inversion schemes can be represented by linear operators both for interferometric and deflectometric cases [49], [50]. The final result at the end of all such reconstruction procedures is an estimate of the instantaneous density uncertainty for each grid point.

In the following sections the uncertainty quantification methodology is tested with synthetic BOS images of a Gaussian density field to assess the performance of the various PIV displacement uncertainty schemes and the propagation framework. Subsequently, the potential of the method is demonstrated with experimental BOS images for supersonic flow over a wedge.

## 3. Analysis with synthetic images

The error analysis is performed using synthetic BOS images rendered using a ray-tracing based image generation methodology, where light rays emerging from the dot pattern are traced through the density gradient field and the optical components of the experimental setup, up to the camera sensor to render the final image. This methodology has been validated using analytical solutions for known density field and the rendered images display realistic features of typical BOS experimental setups such as optical aberrations and blurring due to non-linearities in the density field [51].

The density field chosen for the error analysis is a Gaussian density field, described by Equation (10),

$$\rho(X,Y) = \rho_0 + \Delta\rho_0 \exp\left\{-\frac{(X-X_0)^2 + (Y-Y_0)^2}{2\sigma_0^2}\right\} \tag{10}$$

where $\rho_0$ is the ambient density, $\Delta\rho_0$ is the peak density difference and $\sigma_0$ is the standard deviation of the Gaussian field. This field was chosen because it contains significant displacement gradients to test the displacement uncertainty schemes and the density integration procedure. For the simulations reported in this paper, $\rho_0$ was set to be 1.225 kg/m³, $\Delta\rho_0$ was set to be 0.3 kg/m³, and $\sigma_0$ was set to be 1/4th of the field of view (= 2.41 mm). The dimensions of the density gradient field were 10 x 10 x 10 mm, and it was located at a distance of 0.25 m from the dot pattern. The optical layout used to image the dot pattern and the density field consisted of a 105 mm lens at a distance of 0.5 m from the dot pattern to provide a magnification of about 40 μm/pix. A 2D slice of the three-dimensional density field is shown in Figure 2 (a), and the corresponding light ray displacements are shown in Figure 2 (b). A three-dimensional volume was created using the same slice stacked along the Z direction (out of plane) to account for the depth averaging limitation of BOS experiments.

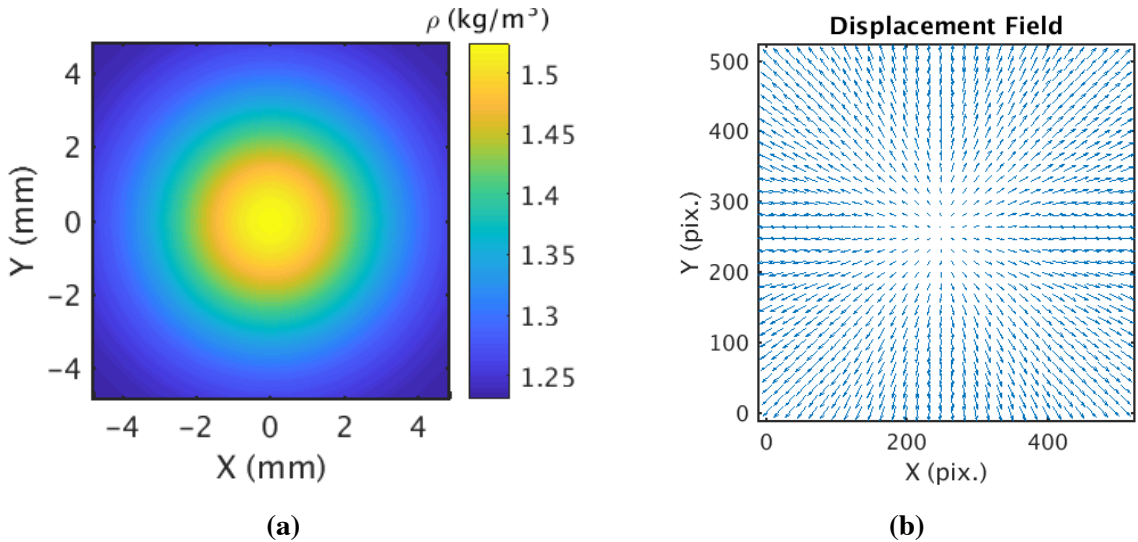

(a)  (b)

Figure 2. (a) 2D slice of the density field used to render the synthetic BOS images, and (b) the corresponding displacement field.

The images were rendered with a dot size of 3 pix. under diffraction limited imaging, with about 20 dots per 32x32 window. The rendered images were corrupted with noise drawn randomly from a zero-mean Gaussian distribution with a standard deviation of 5% of the peak image intensity. A thousand image pairs were rendered in total to create sufficient statistics for the analysis.

The images were processed using a standard cross-correlation procedure for two passes in an iterative window deformation framework [34], [35] with continuous window offset [52]–[54]. The window resolution was 32x32 pix for both passes, which corresponds to a 64x64 pix window size and apodized using a 50% Gaussian window, to minimize edge discontinuities, spectral leakage and wraparound aliasing [55]. The window overlap was set to 0% (grid resolution = 32x32 pix.) for the analysis to avoid introducing covariance on adjacent displacement vectors from the cross-correlation process, as accounting for this covariance in an automatic calibration-free manner is still a subject of ongoing research [40]. The results of the first pass were validated using the Universal Outlier Detection (UOD) method [56] and smoothed, while the results of the second pass were not validated. The displacement uncertainties were calculated using the Image Matching (IM), Correlation Statistics (CS) and Moment of Correlation (MC) methods. For the IM and MC methods, the processing and uncertainty calculation was performed using an open source code PRANA (https://github.com/aether-lab/prana/). For CS, the processing and uncertainty calculation was performed with DaVis 10.0.5 by LaVision. A sample instantaneous displacement field along with the corresponding uncertainty field is shown in Figure 3. Sample instantaneous magnitudes of the displacement and uncertainty fields for (a) Prana, IM, MC and (b) DaVis, CS.(a) for results from PRANA processing and uncertainties from IM and MC, and in Figure 3. Sample instantaneous magnitudes of the displacement and uncertainty fields for (a) Prana, IM, MC and (b) DaVis, CS.(b) for results from DaVis processing and uncertainties from CS.

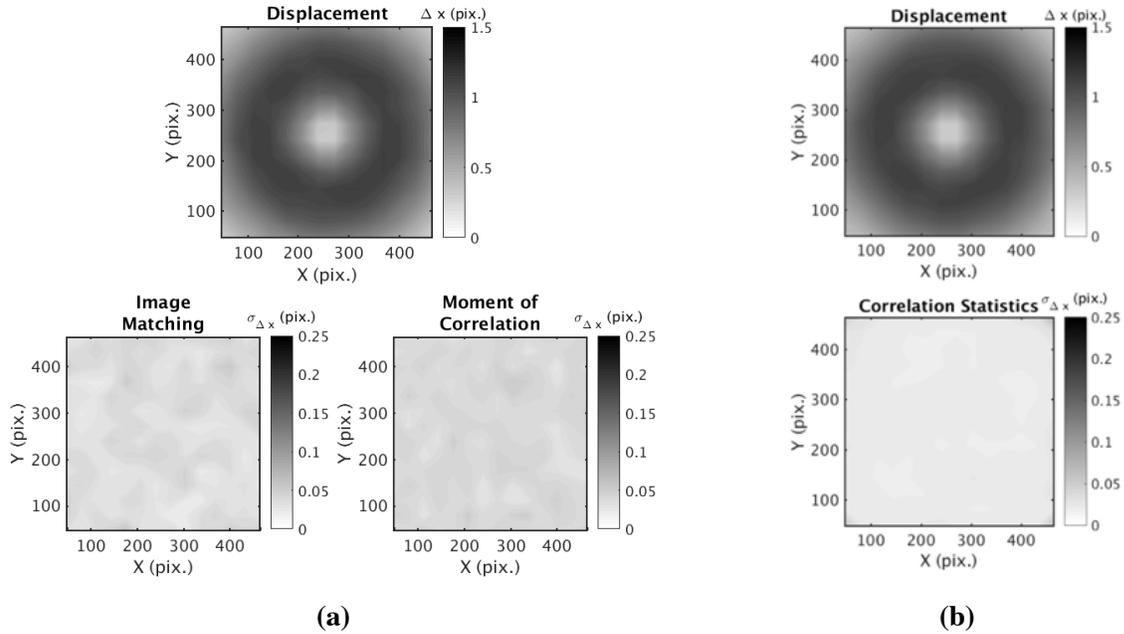

**Figure 3.** Sample instantaneous magnitudes of the displacement and uncertainty fields for (a) Prana, IM, MC and (b) DaVis, CS.

For the error analysis, the displacements obtained from the cross-correlation analysis were compared with the light ray displacements from the ray tracing based image generation procedure to calculate an error for each vector. As the final locations of the light rays will be randomly scattered on the image sensor, an interpolation procedure was performed using a natural neighbor interpolation based on Voronoi tessellations [57] to calculate the corresponding true displacement for each vector. Finally, the errors corresponding to all vectors from all image pairs were combined to build a probability density function (PDF) for the error distribution and the corresponding error statistics such as the bias error, the random error and the total error were calculated. As each image pair yielded 256 vectors with the above processing procedure, with a total of 1000 images, we have 256000 vectors to calculate the statistics. The error statistics were split into three main components, the bias/systematic error, the random error, and the total error, defined as:

$$\begin{aligned}\delta_{bias} &= E(u - u_{true}), \\ \delta_{random} &= \sqrt{E((u - u_{mean})^2)}, \\ \delta_{total} &= \sqrt{E((u - u_{true})^2)} = \sqrt{\delta_{bias}^2 + \delta_{random}^2},\end{aligned}$$ (11)

where $\delta$ represents the error statistic, $u$ is the measurement, $u_{true}$ is the ground truth, and $u_{mean}$ is the average of the measurements.

The displacement uncertainty estimates from the three direct schemes were compared to the random error from the analysis to assess the performance of these PIV-based schemes for synthetic BOS images. The spatial distribution of the error statistics as well as the displacement uncertainties are shown in Figure 4 (a) for PRANA-IM-MC processing and Figure 4 (b) for DaVis-CS processing. It can be seen that the error statistics are fairly uniform throughout the field of view, with negligible bias error from both processing software programs, and that DaVis results in a slightly higher error near the center of the FOV. For the spatial variation of the displacement uncertainty, all three uncertainty schemes result in nearly uniform uncertainty estimates throughout the field of view, and on the same order of their respective random errors. It is important to note that the MC uncertainties reported in this paper are without the bias term in contrast to the original formulation proposed by Bhattacharya et. al. [33]. This is because the bias term in the method is based on an estimate of the local displacement at the end of a converged deformation process, and since the displacement estimation is itself random, the bias estimation itself becomes a random process.

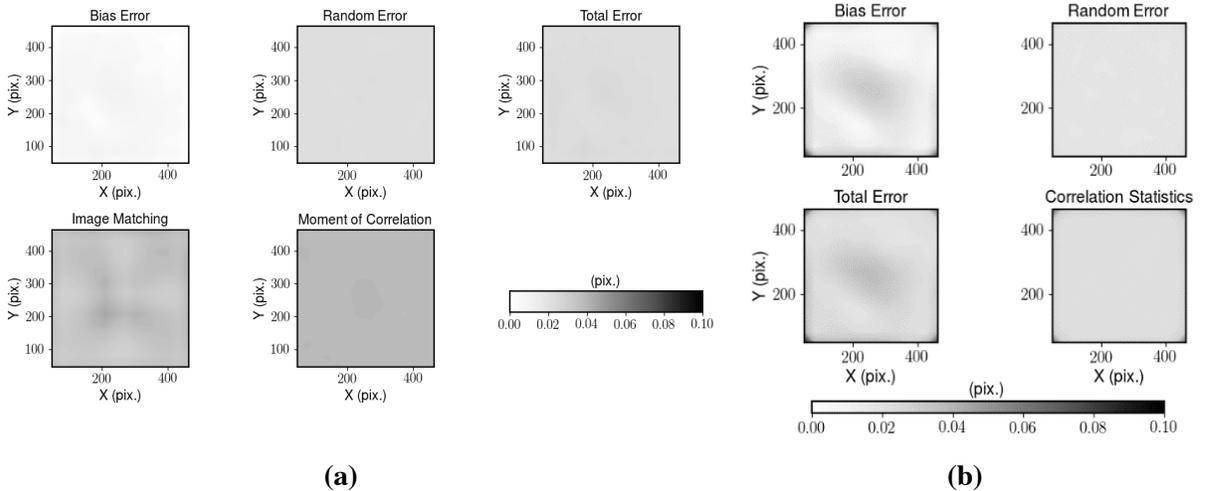

(a)  (b)

**Figure 4.** Spatial variation of the magnitudes of the displacement error statistics and ensemble averaged uncertainty fields for (a) PRANA, IM and MC, and (b) DaVis and CS methods.

The PDFs of the errors and uncertainties from both software programs are shown in Figure 5, along with dashed lines indicating the RMS values of the random error and the RMS values of the corresponding uncertainty schemes. The PDFs were calculated by combining the x and y components of the displacements into a single array. The results for PRANA-IM-MC are shown in Figure 5 (a), and the results for DaVis-CS are shown in Figure 5 (b). It is expected that for a correct uncertainty prediction, the RMS of the random error should coincide with the RMS of the uncertainty distribution [25]. From the figures it can be first seen that all three displacement uncertainty schemes overpredict the corresponding random error, but the RMS of the uncertainty from CS is closest to the RMS of the random error in Figure 5 (b), followed by IM and then MC. Further, CS has a very narrow distribution of the uncertainties compared to IM and MC. The error and uncertainty statistics are summarized in Table 1.

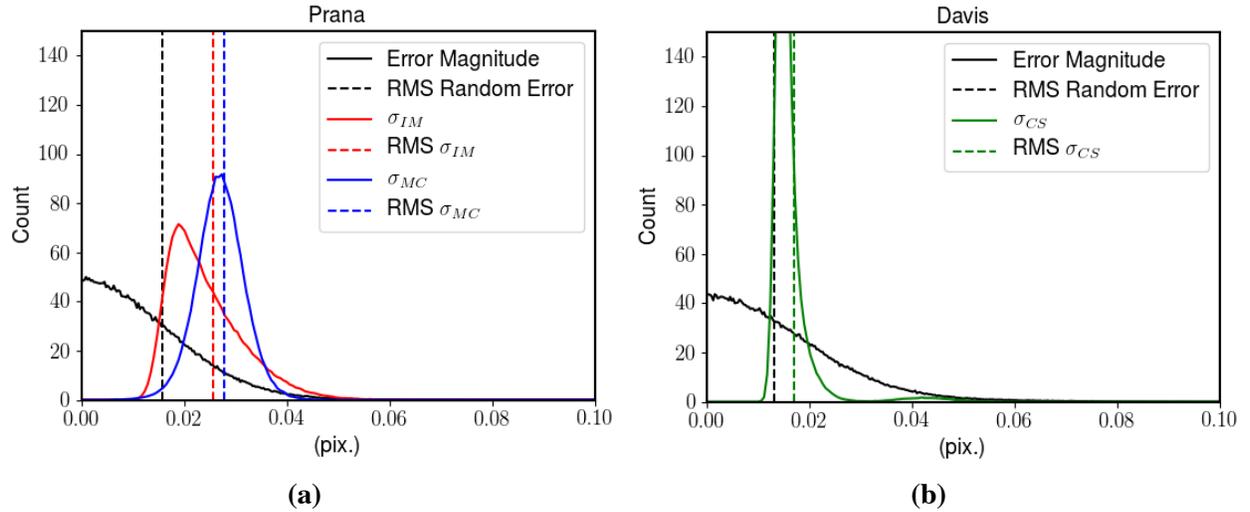

(a)                                    (b)

**Figure 5.** Probability density functions (PDF) of the displacement error and uncertainty distributions along with the corresponding RMS values. (a) PRANA, IM and MC, (b) DaVis, CS

**Table 1.** Displacement error and uncertainty statistics from the two software programs and three uncertainty schemes for the Gaussian density field. All values in units of $\boldsymbol{pix}$.

| PRANA | | DaVis | |
|---|---|---|---|
| Bias Error | 4.45e-03 | Bias Error | 1.45e-02 |
| **Random Error** | **1.58e-02** | **Random Error** | **1.31e-02** |
| Total Error | 1.64e-02 | Total Error | 1.95e-02 |
| Image Matching | 2.56e-02 | Correlation Statistics | 1.70e-02 |
| Moment of Correlation | 2.77e-02 | | |

The displacement fields were also used to calculate the projected density gradient fields using Equation (1), and spatially integrated using the Poisson solver to obtain the projected density field. The thickness of the density gradients from the simulation was then used to calculate the depth averaged density field. Dirichlet boundary conditions were imposed for the density integration procedure, and the density at all four boundaries was set to be values from the true density field used to render the images. Sample results of the depth-averaged density gradient and density fields are shown in Figure 6(a).

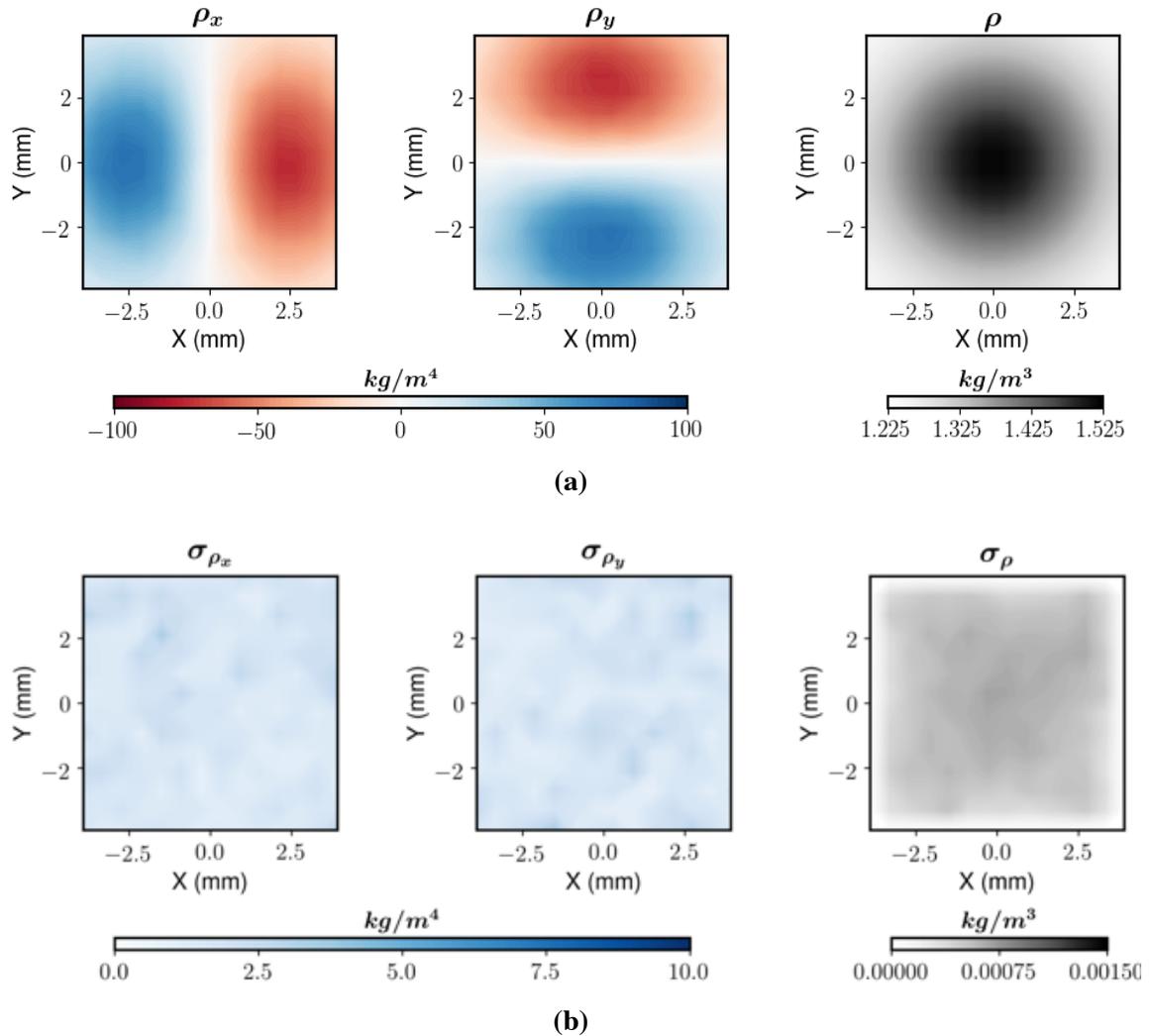

**Figure 6.** Sample instantaneous depth-averaged (a) density gradients and density fields and (b) associated uncertainties obtained from the Poisson solver for PRANA processing.

In addition, the displacement uncertainties from the cross-correlation analysis from each scheme were propagated through the Poisson solver to calculate the density uncertainties. Dirichlet boundary conditions were also used for the uncertainty propagation procedure, with the boundary uncertainties on the four sides set to be 0. Sample instantaneous uncertainties in the depth-averaged density gradient and density fields are shown in Figure 6(b).

The calculated density field was then compared with the original density field used to render the synthetic images and the density error was calculated. The resulting density errors from all 1000 images were used to calculate error statistics. The density error statistics and the corresponding ensemble averaged density uncertainties are shown for results from Prana-IM-MC in Figure 7(a), and for results from DaVis-CS in Figure 7(b).

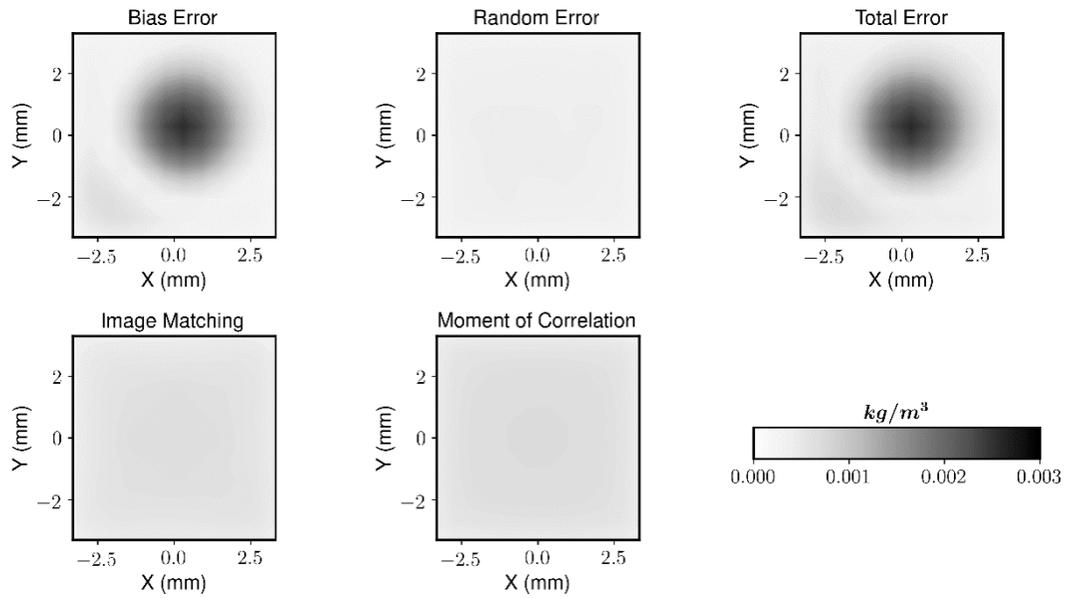

(a)

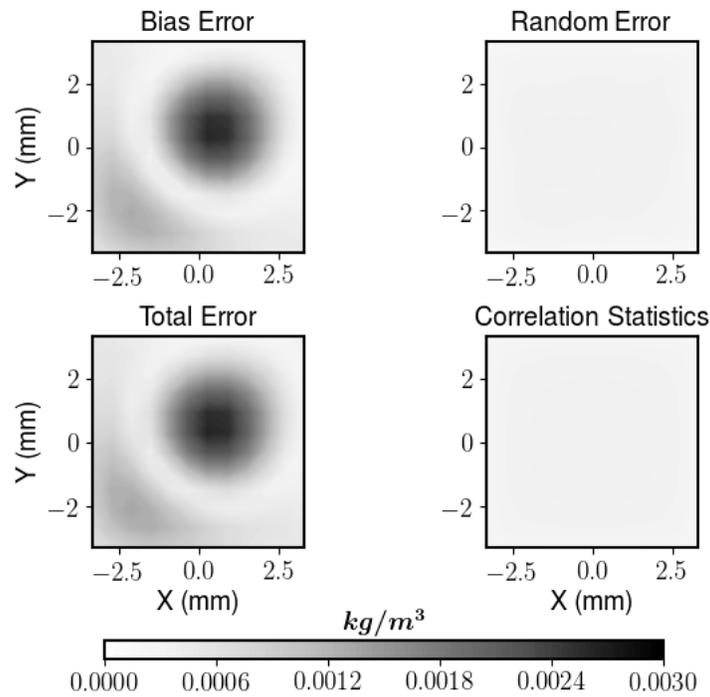

(b)

Figure 7. Spatial variation of density error and uncertainty statistics from (a) PRANA, IM and MC, and (b) DaVis and CS.

From the figures it can be seen that unlike the displacement error statistics, the density error statistics show a strong bias error component (~ 2e-3 kg/m³) as compared to the random error. However, since the uncertainty estimated using the proposed methodology is the random uncertainty, the comparison will be performed between the random error and the uncertainty prediction. The density uncertainty predictions however are spatially uniform for all three methods similar to the displacement uncertainty results shown in Figure 4, and on the same order as the random error (~ 5e-4 kg/m³). But overall, the density uncertainties are seen to be very small likely due to the smoothing nature of the Poisson solver and the uncertainty being zero at the boundaries.

The PDFs of the density error and uncertainty distributions are shown in Figure 8 (a) for PRANA-IM-MC and in Figure 8 (b) for DaVis-CS, along with the corresponding RMS values. Due to the strong bias error in the density results, the RMS of the *random error* will be compared to the RMS of the density uncertainty distributions, and a closer match signifies a better performance. As in the displacement uncertainty results, it is again seen that CS gives the best match between the RMS of the random error and the uncertainty, followed by IM and MC. It is also interesting to note that unlike the displacement error PDFs, the density error PDFs are non-Gaussian, and skewed towards the negative values, signifying that the density error is primarily due to under-prediction. The skewness of the error distribution is also consistent with the strong bias error seen in the spatial error maps in Figure 7. The density errors and uncertainties are summarized in Table 2.

Overall, it is seen from the analysis that (1) PIV-based direct displacement uncertainty schemes are also applicable for BOS images, and (2) Correlation Statistics (CS) performs the best in both the displacement and density uncertainty prediction, though the density results showed a strong anisotropic bias error.

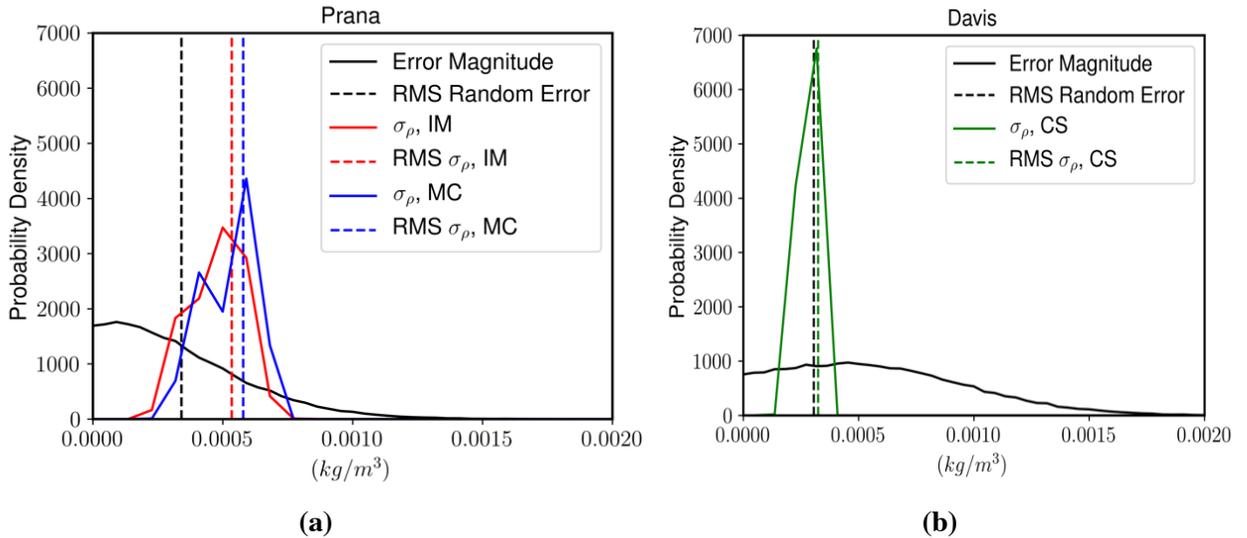

**Figure 8.** PDFs of the density error and uncertainty distributions for (a) PRANA error, IM, and MC uncertainty, and (b) DaVis error and CS uncertainty.

**Table 2.** Density error and uncertainty statistics from the two software programs and three uncertainty schemes for the Gaussian density field. All values in units of $kg/m^3$.

| PRANA | | DaVis | |
|---|---|---|---|
| Bias Error | 8.65e-04 | Bias Error | 9.83e-04 |
| **Random Error** | **3.41e-04** | **Random Error** | **3.06e-04** |
| Total Error | 9.12e-04 | Total Error | 1.02e-03 |
| Image Matching | 5.34e-04 | Correlation Statistics | 3.25e-04 |
| Moment of Correlation | 5.79e-04 | | |

## 4. Demonstration with Experimental Images

The feasibility of the proposed uncertainty quantification methodology is demonstrated with experimental BOS images taken in a supersonic wind tunnel for Mach 2.5 flow over a 11.5° wedge with a base of 1 cm and a height of 2.5 cm. The dot pattern consisted of 0.15 mm diameter dots (corresponding to an image diameter of about 4 pix.) randomly distributed on a transparency with about 25 dots per 32x32 pix. window, and was back-illuminated using an LED with a diffuser plate to obtain uniform illumination. The dot pattern was imaged through the flow with a Photron SAZ camera and a Nikon 105 mm lens at a magnification of 50 um/pix. and an f-number of 32. A total of 5000 images were acquired at 3 kHz for a total/stagnation pressure of 70 psia, corresponding to a free-stream density of 0.49 $kg/m^3$. A layout of the experimental setup is shown in Figure 9 (a), and the wedge geometry is shown in Figure 9 (b).

To account for the startup transients in the tunnel, the images are only recorded during the steady state operation of the tunnel. Further, to avoid masking based errors from affecting the analysis, only a small portion from the flow beneath the wedge is considered in this analysis, and a sample image of the dot pattern with the region of interest (ROI) is shown in Figure 9 (c).

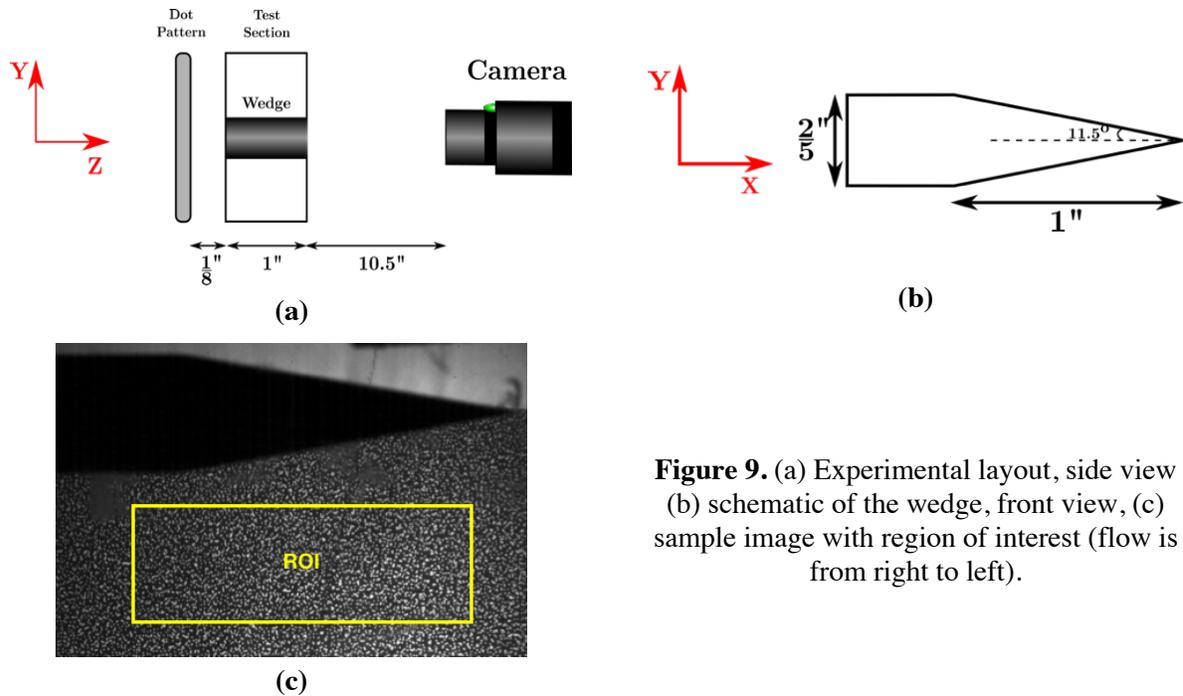

**Figure 9.** (a) Experimental layout, side view (b) schematic of the wedge, front view, (c) sample image with region of interest (flow is from right to left).

The images were processed using the multi-pass window deformation approach described in the previous section for three passes with identical window sizes and overlap percentages (32x32 pix window size and 0% window overlap), with the intermediate pass results smoothed, but without any outlier detection. This was done to preserve the sharp change in displacement in the shock regions, and to prevent them from being identified as be an outlier. The images were processed using PRANA with displacement uncertainty calculation from IM and MC, and using DaVis with displacement uncertainty calculation from CS. To reduce the effect of tunnel/camera/dot-pattern vibrations on subsequent calculations, the displacements in the FOV were subtracted by the average displacements measured in the free-stream region. This was done because the free-stream region ahead of the shock does not contain any density gradients and hence any displacements in this region would be a result of vibrations. To reduce the effect of stray displacements on the density integrations, vectors with displacements less than 0.1 pixels were set to zero throughout the ROI along with the corresponding displacement uncertainties.

The filtered displacements were then used to calculate the density gradients and density fields using the Poisson solver previously described. For the density integration, Dirichlet boundary conditions were used on the right boundary, where the density was set to be its free-stream value of 0.49 kg/m³ and Neumann boundary conditions were imposed on the other three boundaries. Sample images of the path averaged density gradients and density fields are shown in Figure 10 (a) for PRANA processing, and it can be seen the gradients are highest in the regions corresponding to the shock and expansion fan. The density is seen to increase across the shock followed by a decrease across the expansion fan.

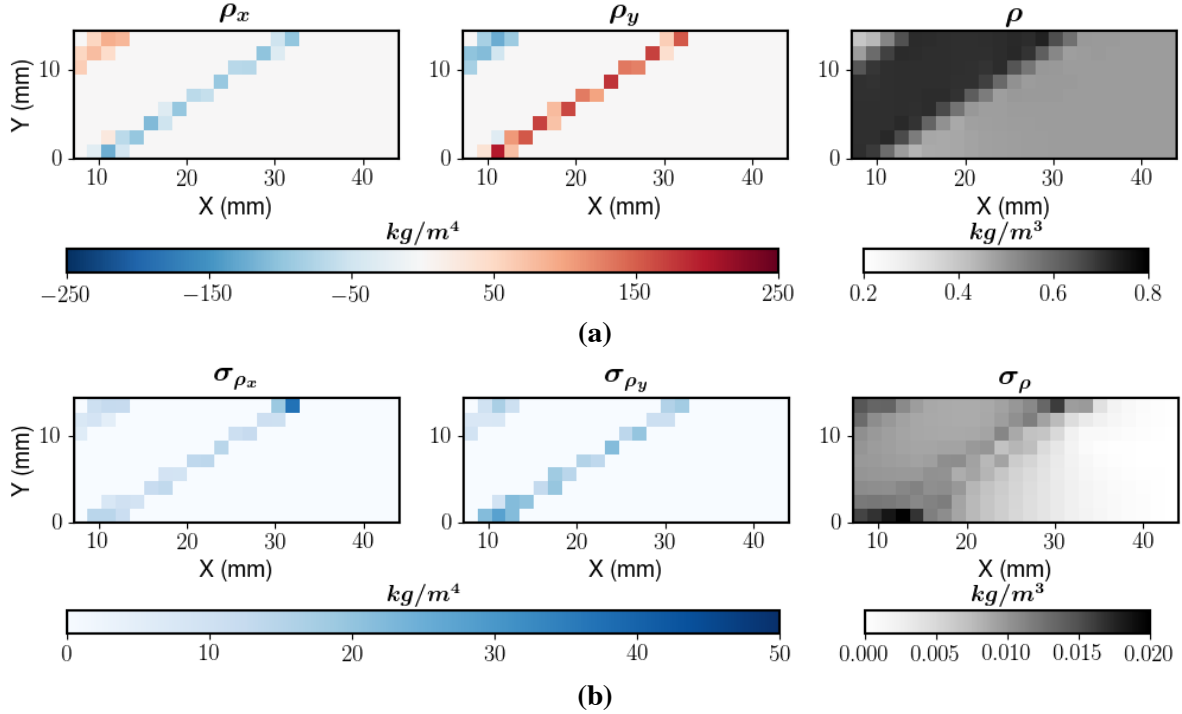

**Figure 10.** (a) Sample density gradient and density fields from PRANA processing and (b) associated uncertainty fields from Image Matching.

A similar approach was also followed for the uncertainty propagation, where a Dirichlet boundary condition was used on the right boundary with an uncertainty of 0.0 kg/m³ as any fluctuations in the free-stream were much lower than the uncertainties measured from BOS, and Neumann boundary conditions were imposed on the other three sides with the measured density gradients. Sample instantaneous uncertainties in the density gradient and density fields are shown in Figure 10 (b) where it seen that the highest uncertainties in the density gradients are also observed in the regions with shocks and expansion fans which correspond to the highest density gradients. However, it is also seen that the region aft (to the left) of the shock has a higher density uncertainty than the region before (to the right of) the shock, even though these points had identical density gradient uncertainties. Further the density uncertainty field is seen to monotonically increase from the right to left for nearly all data points. This exemplifies the uncertainty propagation characteristics of the Poisson solver used for density integration. As the boundary conditions are only specified on the right boundary, the number of points that affect the density estimation at a given point increases as one moves to the left, and hence the density uncertainty at the given point is also a combination of the uncertainties from an increasing number of points. Since the density gradient uncertainty is always positive, the result is that the density uncertainty field monotonically increases. As a consequence, this can be generalized to the statement that the density uncertainty for a BOS experiment will always increase as one moves away from the Dirichlet boundaries. This is an artifact of the density integration procedure using the Poisson equation, and represents one of the method's limitations.

The uncertainty fields across five thousand images were averaged to calculate the statistics, and the ensemble averaged field is shown in Figure 11 for all three uncertainty schemes. Qualitatively, it is seen that the ensemble averaged uncertainty distributions are very similar to the instantaneous fields shown in Figure 10 (b), with the highest uncertainties in the shock region and a monotonic increase of uncertainty from right to left. It is also seen that while IM and CS result in similar density uncertainty, the MC method under-predicts in comparison to the other two schemes.

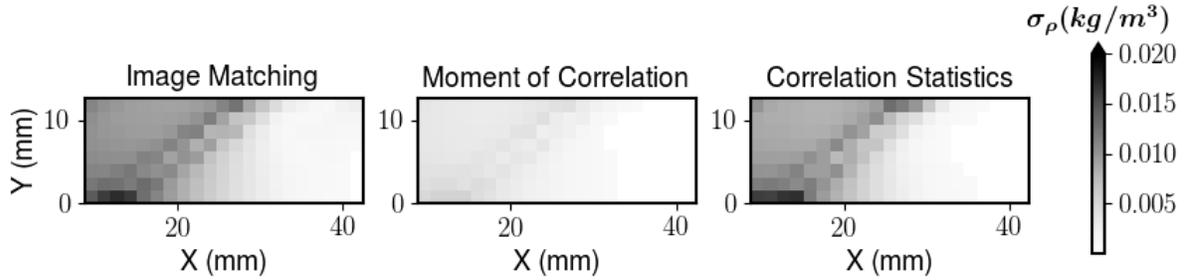

**Figure 11.** Spatial variation of ensemble averaged density uncertainty predictions from Image Matching, Moment of Correlation and Correlation Statistics schemes.

Finally, the uncertainties from all vectors in the time series are combined to calculate the PDFs for the uncertainty distributions. The resulting PDFs are shown in Figure 12 along with the RMS values which are 9.34e-3 kg/m³ for IM, 3.48e-3 kg/m³ for MC and 8.51e-3 kg/m³ for CS., It is seen that MC results in the lowest uncertainty and IM results in the highest uncertainty with CS predicting a value slightly lower than IM.

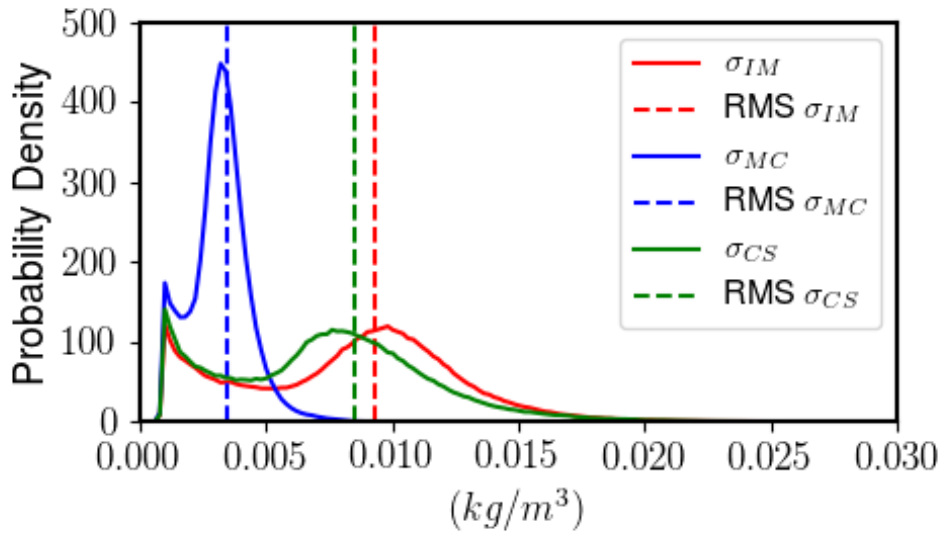

**Figure 12.** PDFs of the density uncertainty distributions for IM, MC, and CS.

## 5. Conclusion

We have implemented and presented the first comprehensive uncertainty quantification framework for density estimation from BOS measurements and tested the method with synthetic and experimental BOS images. The methodology builds upon recent progress in a-posteriori uncertainty quantification in PIV, and direct displacement uncertainty methods are used to also estimate the displacement uncertainty from BOS images. These displacement uncertainties are then propagated to the density gradients using the optical layout and then through the Poisson solver typically used for density integration in BOS to calculate density uncertainties, accounting for the covariances introduced due to the finite differences involved in the calculation of the Laplacian. This method yields instantaneous, local uncertainty bounds for each density measurement throughout the field of view.

The methodology was tested with synthetic BOS images rendered with a Gaussian density field using a ray-tracing based image generation methodology. The images were processed using correlation algorithms with multi-pass window deformation, and the errors were calculated by comparing the measured displacements to the light ray displacements, which are considered to be the ground truth. Processing was done using two different software programs, PRANA and DaVis, and three displacement uncertainty estimation schemes– Image Matching (IM), Moment of Correlation (MC) and Correlation Statistics (CS). Results show that for the displacements, all methods overpredict the true random error, with CS closest to the random error, followed by IM and MC.

When propagated through the Poisson solver for density integration, results from both processing software programs resulted in a stronger bias error in the density field, likely due to truncation errors from the finite differences used in the density integration process. On comparing the random errors with the predicted density uncertainties, CS predicted a density uncertainty closest to the corresponding random error. IM and MC both overpredicted their respective random errors, but IM was closer to the true random error compared to MC.

The method was also demonstrated on experimental BOS images of supersonic flow over a wedge and the processed displacements and the density fields show the presence of a shock wave and expansion fan in the region of interest corresponding to the wedge tip and wedge shoulder respectively. The density gradient uncertainties were highest in close proximity to the shocks and expansion fans, and were sensitive to the boundary condition and the integration procedure. In general, the density uncertainty increased monotonically on moving away from the Dirichlet boundary, with the result that a point downstream of the shock had a higher density uncertainty as compared to a point upstream of the shock, even though they had nearly identical density gradient uncertainties. PDFs of the uncertainty fields from five thousand vector fields showed that IM and CS resulted in very similar uncertainties, and MC under-predicted the uncertainty as compared to the two methods.

A limitation of the proposed methodology is that bias uncertainties are not estimated, and as seen from the analysis with the synthetic images, there is a strong bias error in the density estimation. This is in addition to the bias uncertainties that also exist in the cross-correlation processing which are due to peak-locking and other processing based errors. Developing a similar formulation for the estimation and propagation of the bias uncertainties is still required. Another limitation is that the methodology does not account for covariances introduced between adjacent vectors due to the cross-correlation procedure. This is especially important in situations where window overlap is used in the processing, and is another avenue for future work on this topic. Finally, further work is required to compare these density uncertainty predictions to the measurement error for benchmark BOS experiments.

Overall, displacement uncertainty methods typically used for PIV experiments are also applicable to BOS data, and the displacement uncertainties can be propagated through the Poisson solver using a sparse linear operator to obtain the density uncertainties. Thus, the method proposed in this manuscript allows for instantaneous spatially resolved uncertainty quantification in density estimates from BOS measurements, and for use in CFD model validation and engineering design.

## Acknowledgements

This material is based upon work supported by the U.S. Department of Energy, Office of Science, Office of Fusion Energy Sciences under Award Number DE-SC0018156.